\documentclass[reprint, superscriptaddress,  aps]{revtex4-2}
\usepackage{amssymb}
\usepackage{amsmath}
\usepackage{graphics}
\usepackage{graphicx}
\usepackage{subfigure}
\usepackage{dcolumn}
\usepackage{bm}
\usepackage{hyperref}
\usepackage{color}
\usepackage{epstopdf}

\begin{document}

\title{Efficient high-harmonic generation in graphene with two-color laser
field at orthogonal polarization}

\author{Hamlet K. Avetissian}
\affiliation{Centre of Strong Fields Physics, Yerevan State University,
Yerevan 0025, Armenia}

\author{G.F. Mkrtchian}
\thanks{mkrtchian@ysu.am}
\affiliation{Centre of Strong Fields Physics, Yerevan State University,
Yerevan 0025, Armenia}

\author{Andreas Knorr}
\affiliation{Nichtlineare Optik und Quantenelektronik, Technische
Universit\"{a}t Berlin, Berlin 10623, Germany}

\begin{abstract}
High-order frequency mixing in graphene using a two-color radiation field
consisting of the fundamental and the second harmonic fields of an
ultrashort linearly polarized laser pulse is studied. It is shown that the
harmonics originated from the interband transitions are efficiently
generated in the case of the orthogonally polarized two-color field. In this
case, the generated high-harmonics are stronger than those obtained in the
parallel polarization case by more than two orders of magnitude. This is in
sharp contrast with the atomic and semiconductor systems, where parallel
polarization case is more preferable. The physical origin of this
enhancement is also deduced from the three-step semi-classical electron-hole
collision model, extended to graphene with pseudo-relativistic energy
dispersion. In particular, we discuss the influence of the many particle Coulomb
interaction on the HHG process within dynamical Hartree-Fock approximation.
Our analysis shows that in all cases we have an overall enhancement of the
HHG signal compared with the free-charged carrier model due to the
electron-hole attractive interaction.
\end{abstract}

\maketitle

\section{ Introduction}

Significant experimental and theoretical efforts have recently been invested
in the generation of high-order harmonics (HHG) of a laser field interacting
with the condensed phase of the matter. The generation of harmonics has recently
been reported in solid-state materials with crystalline symmetry \cite%
{Ghimire11,Zaks,Schubert,Vampa2015,Luu,Hohenleutner,You,Langer,Ghimire,Uzan,Bowlan,Yoshikawa,Hafez,Giorgianni,Breton,Liu}%
, in amorphous solids \cite{You17}, and in liquids \cite{Luu2018}, which
ensures that the condensed phase of the matter has the potential of becoming
a compact and efficient attosecond light source of the next-generation due
to the high density of the emitters, compared with a gaseous medium. 
HHG gives access to a frequency range that is difficult to
achieve in other ways \cite{Avetissian-book}, and provides a
frequency-domain view of the electron dynamics in quantum systems \cite%
{Smirnova}. The complete characterization of the radiated harmonic spectrum,
phase, and polarization will allow to recover the underlying
quasiparticle dynamics in solids. In particular, using HHG one can
reconstruct the crystal potential and electrons density with a spatial
resolution of about 10 picometres \cite{Lakhotia}, which can be used for the
direct investigation of the electronic and topological properties of the
materials. From the spectra of HHG in crystals one can observe the dynamical
Bloch oscillations \cite{Luu}, Mott \cite{Silva}, and Peierls \cite{Bauer}
transitions, retrieve the band-structure \cite{Vampa2,Dejean} or the band
topology \cite{Luu218}.

The HHG in atomic gases has been intensively investigated since the 90s of
the last century \cite{Corkum93} and now with the advent of controlled
few-cycle light waves is the basis of the attosecond science \cite%
{Corkum2007,Ivanov2009}. In view of the vast theoretical and experimental
methods developed for atomic HHG, it is of interest to analyze whether
methods that were developed for enhancing HHG in the gas phase are also
applicable to HHG in solid state nanostructures. Among the existing nanostructures, the
graphene due to its more pronounced properties allows to use it as a more
effective nonlinear optical material and has triggered many studies devoted
to nonperturbative HHG \cite%
{g1,g2,g3,g4,g5,g6,g7,g8,g9,g10,Mer18,g11,go12,go13,go14,go15}.\textrm{\ }%
Specifically, diverse polarization and ellipticity dependence effects in the
total HHG spectrum \cite{go12,go13,go14,go15} are revealed in a monolayer
graphene where interaction with the pump wave drives charged carriers far
away from the Dirac cones. The latter opens up wide opportunities for
increasing the HHG yield in graphene by choosing the parameters of the
driving wave.

An efficient option for increasing the HHG yield in atomic systems is the
use of multicolor driving pulses. High-order wave mixing processes with
multi-color laser fields have gained enormous interest due to the additional
degrees of freedom, such as relative polarizations, intensities, phases, and
wavelengths of involved pump waves. In atomic systems, it has been shown
that adding an additional second or third harmonic field can enhance the HHG
in the plateau region by several orders of magnitude \cite{Atom1,Atom2,Atom3}%
. The HHG with different compositions of the driving laser pulses was
addressed also for solid targets and nanostructures considering two distinct
regimes: First, if the driving field consists of the fundamental wave and
its harmonics \cite%
{Vampa2015,Vampa2,Li,Worner,Shirai,Song,Shao,Navarrete,Mrudul}, and second,
if one of the involved wave frequencies significantly higher than the other
one \cite{Zaks,Yan,Crosse,Xie,Langer16,Langer18,Mer19-2,Mer20}. Two-color
high-order wave mixing research reported so far has mainly been performed
for gapped systems. For 2D semimetals, two-color high-order wave mixing was
considered in Ref. \cite{Mer19-2} in case when one of the frequencies
significantly higher than the other one. In Ref. \cite{Mrudul} it was
considered valley-selective HHG in pristine graphene by using a combination
of the two counter-rotating circularly polarized fields. For solid targets
with an energy gap exposed to two- or three-color laser pulses \cite%
{Li,Song,Navarrete} of parallel polarizations the enhancement of HHG has
been shown in an analogy with the atomic HHG \cite{Atom1}, where the HHG in
the orthogonally polarized two-color field is suppressed. The latter is
intuitively clear in terms of a simple quasiclassical three-step model \cite%
{Corkum2007,Ivanov2009}. As far as the polarization of resultant field is no
longer linear, the ionized electron will, in general, never return to the
parent ion. However, for gapless systems, as will be shown in the current
paper, this conclusion does not hold, and the opposite applies: the HHG yield will be
increased considerably at the orthogonally polarized two-color driving wave
fields.

In this paper the high-order frequency mixing in graphene using a two-color
radiation field that consisted of the fundamental and the second harmonic
(SH) fields of an ultrashort linearly polarized laser pulse is studied. It
is shown that the harmonics originated from the interband transitions are
efficiently generated in the case of the orthogonally polarized two-color
laser field. This is in sharp contrast with the atomic \cite{Atom1} and
semiconductor cases \cite{Navarrete} where the parallel polarization case is
more preferable.

We consider a rather general model, including many-particle Coulomb
interaction, since the Coulomb interaction is known to be generally stronger
in low-dimensional structures. The importance of Coulomb interaction for
ultrafast many-particle kinetics in graphene has been theoretically and
experimentally verified \cite{Knorr1,Knorr2,Knorr-book}. The significance of
many-body Coulomb interaction at the high harmonic generation process in
graphene has been shown in Ref. \cite{Mer18}. The latter study was conducted
near the Dirac points and the open question of interest remains: to develop
the theory of HHG with many-body Coulomb interaction beyond the Dirac cone
approximation that is applicable to the full Brillouin zone (BZ) \cite%
{Knorr2}. As a first step, the combined carrier-carrier and carrier-phonon
scatterings are taken into account phenomenologically with relaxation rate
in femtosecond time-scale \cite{Knorr1}.

The paper is organized as follows. In Sec. II the evolutionary equation for
the single-particle density matrix is presented. The electron-electron
Coulomb interaction is taken into account within the dynamical Hartree-Fock
(HF) approximation \cite{Kira,Knorr-book} beyond the Dirac cone
approximation and applicable to the full Brillouin zone of a hexagonal
tight-binding nanostructure. In Sec. III, we consider HHG spectra and
present the main results. Finally, conclusions are given in Sec. IV.

\section{The evolutionary equation for the single-particle density matrix}

We consider the interaction of a strong two-color laser field $\mathbf{E}(t)$
with graphene. The waves propagate in a perpendicular direction to the
monolayer plane ($XY$) of graphene with the electric field strength: 
\begin{equation}
\mathbf{E}\left( t\right) =f_{1}\left( t\right) E_{01}\hat{\mathbf{e}}%
_{1}\cos \left( \omega t\right) +f_{2}\left( t\right) E_{02}\hat{\mathbf{e}}%
_{2}\cos \left( 2\omega t\right) ,  \label{Et}
\end{equation}%
where $E_{01}$ and $E_{02}$ are the amplitudes of the laser pulses, $\omega $
is the fundamental frequency, $\hat{\mathbf{e}}_{1}$ and $\hat{\mathbf{e}}%
_{2}$ are the unit polarization vectors. The envelopes of the two
wave-pulses are described by the sin-squared functions%
\begin{equation}
f_{1,2}(t)=\left\{ 
\begin{array}{cc}
\sin ^{2}\left( \pi t/\tau _{1,2}\right) , & 0\leq t\leq \tau _{1,2}, \\ 
0, & t<0,t>\tau _{1,2},%
\end{array}%
\right.  \label{envelop}
\end{equation}%
where $\tau _{1}$ and $\tau _{2}$ characterize the pulse duration. The
carrier-envelope phases for both wave-pulses are set to zero.

\begin{figure}[tbp]
\includegraphics[width=.48\textwidth]{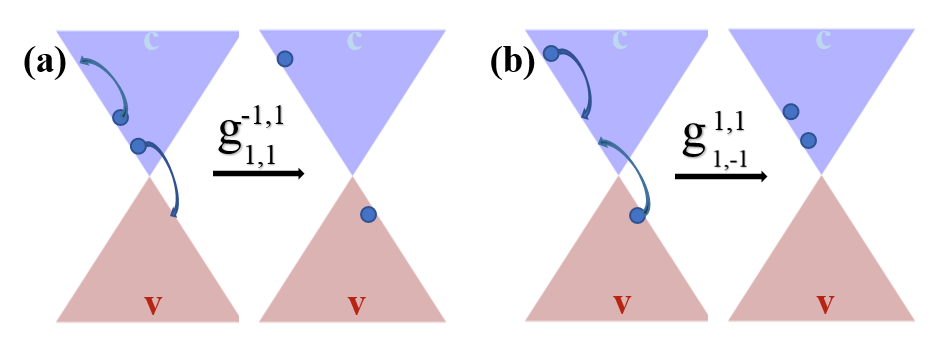}
\caption{Schematic representation of the Coulomb-induced MA processes near
the Dirac cone for destruction (a) and multiplication (b) of carriers in one
specific band at the expense of another band.}
\end{figure}
The dynamics of the system is governed by the total Hamiltonian: 
\begin{equation}
\hat{H}=\hat{H}_{\mathrm{free}}+\hat{H}_{\mathrm{C}}+\hat{H}_{\mathrm{int}},
\label{Ham}
\end{equation}
where 
\begin{equation}
\hat{H}_{\mathrm{\ free}}=\sum_{\lambda ,\mathbf{k}}\mathcal{E}_{\lambda
}\left( \mathbf{k}\right) \hat{e}_{\lambda \mathbf{k}}^{\dagger }\hat{e}
_{\lambda \mathbf{k}}  \label{Hamfree}
\end{equation}
is the free particle Hamiltonian, with $\hat{e}_{\lambda \mathbf{k}}$\ ($
\hat{e}_{\lambda \mathbf{k}}^{\dagger }$) the annihilation (creation)
operators for an electron with the momentum $\mathbf{k}$\ and band index $
\lambda =\pm 1\widehat{=}c,v$ (for conduction ($\lambda =1$) and valence ($
\lambda =-1$) bands). In Eq. (\ref{Hamfree}) $\mathcal{E}_{c}\left( \mathbf{k
}\right) $\ and $\mathcal{E}_{v}\left( \mathbf{k}\right) $ are the
corresponding band energy dispersions. The specific expressions for the
single-particle Hamiltonian and other derived quantities are given in the
Appendix. In HF approximation, we reduce the Coulomb interaction Hamiltonian 
$\hat{H}_{\mathrm{C}}$\ into the mean-field Hamiltonian: 
\begin{equation}
\hat{H}_{\mathrm{C}}=-\ \sum\limits_{\alpha \beta \gamma \delta
}\sum\limits_{\mathbf{k}\neq \mathbf{k}^{\prime }}g_{\gamma ,\beta }^{\alpha
,\delta }\left( \mathbf{k,k}^{\prime }\right) V_{2D}(\mathbf{k}^{\prime }-
\mathbf{k})\rho _{\gamma \delta }\left( \mathbf{k}^{\prime },t\right) \hat{e}
_{\alpha \mathbf{k}}^{\dagger }\hat{e}_{\beta \mathbf{k}}.  \label{Hamc}
\end{equation}
where $V_{2D}\left( \mathbf{k}\right) $\ is the Fourier transform of the
electron-electron interaction potential and the form factor reads:
\begin{equation*}
g_{\gamma ,\beta }^{\alpha ,\delta }\left( \mathbf{k,k}^{\prime }\right) =
\frac{1}{4}\left( 1+\alpha \gamma \delta \beta +\left( \delta \beta +\alpha
\gamma \right) f_{c}\left( \mathbf{k,k}^{\prime }\right) \right.
\end{equation*}
\begin{equation}
\left. +i\left( \delta \beta -\alpha \gamma \right) f_{s}\left( \mathbf{k,k}
^{\prime }\right) \right) .  \label{str}
\end{equation}
In Eq. (\ref{Hamc}) $\rho _{\gamma \delta }\left( \mathbf{k}^{\prime
},t\right) =\left\langle \widehat{e}_{\delta \mathbf{k}^{\prime }}^{\dagger }
\widehat{e}_{\gamma \mathbf{k}^{\prime }}\right\rangle $\ is the single
particle density matrix, and the functions 
\begin{eqnarray*}
f_{c}\left( \mathbf{k,k}^{\prime }\right) &=&\mathrm{Re}\left\{ \frac{
f^{\ast }\left( \mathbf{k}\right) f\left( \mathbf{k}^{\prime }\right) }{
\left\vert f\left( \mathbf{k}\right) \right\vert \left\vert f\left( \mathbf{k
}^{\prime }\right) \right\vert }\right\} , \\
f_{s}\left( \mathbf{k,k}^{\prime }\right) &=&\mathrm{Im}\left\{ \frac{
f^{\ast }\left( \mathbf{k}\right) f\left( \mathbf{k}^{\prime }\right) }{
\left\vert f\left( \mathbf{k}\right) \right\vert \left\vert f\left( \mathbf{k
}^{\prime }\right) \right\vert }\right\}
\end{eqnarray*}
are defined via the structure function $f\left( \mathbf{k}\right) $ (\ref
{f(k)}). The terms in Eq. (\ref{Hamc}) proportional to $f_{s}\left( \mathbf{
k,k}^{\prime }\right) $\ are nonzero when $\delta \beta =-\alpha \gamma $,\
and describe Meitner-Auger processes \cite{Meitner,Auger,MA}. A
Meitner-Auger (MA) process is a Coulomb-induced interaction for which the
number of carriers in the bands are not conserved individually, cp. Fig. 1.
In graphene, due to a vanishing bandgap, these processes cannot be
neglected, in contrast to conventional semiconductors when a large bandgap
suppresses the MA effect by the demand of energy conservation. The two major
MA processes for electrons, corresponding to the Coulomb matrix elements $
g_{1,1}^{-1,1}\left( \mathbf{k,k}^{\prime }\right) $ and $
g_{1,-1}^{1,1}\left( \mathbf{k,k}^{\prime }\right) $ for destruction (a) and
multiplication (b) of carriers in one specific band at the expense of
another band are displayed in Fig. 1. For holes, similar processes occur.

The last term in Eq. (\ref{Ham})
\begin{equation}
\hat{H}_{\mathrm{int}}=ie\mathbf{E}(t)\sum\limits_{\alpha \beta
}\sum\limits_{\mathbf{kk}^{\prime }}\int \left( \partial _{\mathbf{k}
}e^{i\left( \mathbf{k}^{\prime }-\mathbf{k}\right) \mathbf{r}}\right) d
\mathbf{r}\langle \alpha \mathbf{k}||\beta \mathbf{k}^{\prime }\rangle \ 
\hat{e}_{\alpha \mathbf{k}}^{\dagger }\hat{e}_{\beta \mathbf{k}^{\prime }}
\label{intH}
\end{equation}
is the the light-matter interaction Hamiltonian in the length gauge, $e$\ is
the elementary charge, and
\begin{equation*}
\langle \alpha \mathbf{k}||\beta \mathbf{k}^{\prime }\rangle =\frac{1}{2}
\left( f_{c}\left( \mathbf{k,k}^{\prime }\right) +if_{s}\left( \mathbf{k,k}
^{\prime }\right) +\alpha \beta \right) .
\end{equation*}
Now, from the Heisenberg equation $i\hbar \partial _{t}\hat{e}_{\alpha 
\mathbf{k}}^{\dagger }\hat{e}_{\beta \mathbf{k}}=\left[ \hat{e}_{\alpha 
\mathbf{k}}^{\dagger }\hat{e}_{\beta \mathbf{k}},\hat{H}\right] $\ one can
obtain the closed set of evolutionary equations for a density matrix $\rho
_{\beta ,\alpha }(k,t)=\left\langle \hat{e}_{\alpha \mathbf{k}}^{\dagger }
\hat{e}_{\beta \mathbf{k}}\right\rangle $. 
The diagonal elements represent particle distribution functions for
conduction $\mathcal{N}_{c}(\mathbf{k},t)=\rho _{1,1}(\mathbf{k},t)$ and
valence $\mathcal{N}_{v}(\mathbf{k},t)=\rho _{-1,-1}(\mathbf{k},t)$ bands,
and the non-diagonal element is the interband polarization $\mathcal{P}(%
\mathbf{k},t)=\rho _{1,-1}(\mathbf{k},t)$. On the HF level for an undoped
system in equilibrium, the initial conditions $\mathcal{P}(\mathbf{k},0)=0$, 
$\mathcal{N}_{c}^{0}(\mathbf{k})=0$, and $\mathcal{N}_{v}^{0}(\mathbf{k})=1$
are assumed, neglecting thermal occupations or doping. Since $\mathcal{N}_{v}(\mathbf{k%
},t)=1-$ $\mathcal{N}_{c}(\mathbf{k},t)$, the equation for $\mathcal{N}_{v}(%
\mathbf{k},t)$ is superficial. Thus, the graphene interaction with a strong
laser field is modeled as:%
\begin{equation*}
\partial _{t}\mathcal{N}_{c}(\mathbf{k},t)-\hbar ^{-1}e\mathbf{E}\left(
t\right) \partial _{\mathbf{k}}\mathcal{N}_{c}(\mathbf{k},t)=
\end{equation*}%
\begin{equation}
-2\mathrm{Im}\left\{ \left[ \hbar ^{-1}\mathbf{E}\left( t\right) \mathbf{D}_{%
\mathrm{tr}}\left( \mathbf{k}\right) +\Omega _{c}\left( \mathbf{k},t;%
\mathcal{P},\mathcal{N}_{c}\right) \right] \mathcal{P}^{\ast }(\mathbf{k}%
,t)\right\} ,  \label{B1}
\end{equation}%
\begin{equation*}
\partial _{t}\mathcal{P}(\mathbf{k},t)-\hbar ^{-1}e\mathbf{E}\left( t\right)
\partial _{\mathbf{k}}\mathcal{P}(\mathbf{k},t)=
\end{equation*}%
\begin{equation*}
-i\hbar ^{-1}\left[ \mathcal{E}_{c}\left( \mathbf{k}\right) -\mathcal{E}%
_{v}\left( \mathbf{k}\right) -\Xi _{c}(\mathbf{k},t;\mathcal{P},\mathcal{N}%
_{c})-i\hbar \Gamma \right] \mathcal{P}(\mathbf{k},t)
\end{equation*}%
\begin{equation}
+i\left[ \hbar ^{-1}\mathbf{E}\left( t\right) \mathbf{D}_{\mathrm{tr}}\left( 
\mathbf{k}\right) +\Omega _{c}\left( \mathbf{k},t;\mathcal{P},\mathcal{N}%
_{c}\right) \right] \left[ 1-2\mathcal{N}_{c}(\mathbf{k},t)\right] ,
\label{B3}
\end{equation}%
where $\mathbf{D}_{\mathrm{tr}}\left( \mathbf{k}\right) =-e\langle c,\mathbf{%
k}|i\partial _{\mathbf{k}}|v,\mathbf{k}\rangle $ is the transition dipole
moment, $\Gamma $ is the phenomenological relaxation rate. The many-body
Coulomb interaction renormalizes the light-matter coupling via the internal
dipole field of all generated electron-hole excitations: 
\begin{equation*}
\Omega _{c}\left( \mathbf{k},t;\mathcal{P},\mathcal{N}_{c}\right) =\frac{%
\hbar ^{-1}}{\left( 2\pi \right) ^{2}}\int_{BZ}d\mathbf{k}^{\prime
}V_{2D}\left( \mathbf{k-k}^{\prime }\right)
\end{equation*}%
\begin{equation}
\times \left\{ \mathcal{P}^{\prime }\left( \mathbf{k}^{\prime },t\right)
+if_{c}\left( \mathbf{k,k}^{\prime }\right) \mathcal{P}^{\prime \prime
}\left( \mathbf{k}^{\prime },t\right) -if_{s}\left( \mathbf{k,k}^{\prime
}\right) \mathcal{N}_{c}\left( \mathbf{k}^{\prime },t\right) \right\} ,
\label{Rc}
\end{equation}%
as well as the transition energies:%
\begin{equation*}
\Xi _{c}(\mathbf{k},t;\mathcal{P},\mathcal{N}_{c})=\frac{2}{\left( 2\pi
\right) ^{2}}\int_{BZ}d\mathbf{k}^{\prime }V_{2D}\left( \mathbf{k-k}^{\prime
}\right)
\end{equation*}

\begin{equation}
\times \left\{ f_{c}\left( \mathbf{k,k}^{\prime }\right) \mathcal{N}%
_{c}\left( \mathbf{k}^{\prime },t\right) +f_{s}\left( \mathbf{k,k}^{\prime
}\right) \mathcal{P}^{\prime \prime }\left( \mathbf{k}^{\prime },t\right)
\right\} .  \label{Wc}
\end{equation}%
From Eqs. (\ref{B1}) and (\ref{B3}) one can recover semiconductor Bloch
equations for an ideal 2D semiconductor \cite{Guazzotti} taking $f_{c}\left( \mathbf{k,k}^{\prime }\right) =1$ and $%
f_{s}\left( \mathbf{k,k}^{\prime }\right) =0$. The Coulomb contribution (\ref%
{Wc}) in Eq. (\ref{B3}) describes the repulsive electron-electron
interaction and leads to a renormalization of the single-particle energy $%
\mathcal{E}_{c,v}\left( \mathbf{k}\right) $. Note that the Coulomb-induced
self-energy has been absorbed into the definition of the single-particle
energy and will not be written explicitly hereafter. The Coulomb
contribution Eq. (\ref{Rc}) leads to a renormalization of the Rabi frequency
and accounts for electron-hole attraction. Already in linear spectroscopy 
this term is responsible for the formation of excitons in semiconductors. In graphene, it gives rise to
so-called saddle-point exciton \cite{Yang,Mak} near the van Hove singularity
of graphene BZ.

The optical excitation induces a surface current that can be calculated by
the following formula: 
\begin{equation}
\mathbf{j}\left( t\right) =-2e\sum\limits_{\mathbf{k,}\lambda ,\lambda
^{\prime }}\rho _{\lambda ,\lambda ^{\prime }}(\mathbf{k},t)\langle \lambda
^{\prime },\mathbf{k}|\widehat{\mathbf{v}}\left( \mathbf{k}\right) |\lambda ,%
\mathbf{k}\rangle .  \label{sfc}
\end{equation}%
Here $\widehat{\mathbf{v}}\left( \mathbf{k}\right) $ is the velocity
operator, and factor two takes into account spin degeneracy. Note that since
we are integrating over the entire Brillouin zone, only the spin-degeneracy
has been taken into account.

For the numerical solution of Eqs. (\ref{B1}) and (\ref{B3}), we can make a
change of variables and transform the partial differential equations into
ordinary ones. The new variables are $t$ and $\mathbf{k}_{0}=\mathbf{k}-%
\mathbf{k}_{E}$, where $\mathbf{k}_{E}\left( t\right) =-e/\hbar \int_{0}^{t}%
\mathbf{E}\left( t^{\prime }\right) dt^{\prime }$ is the classical momentum
given by the wave field. The latter can be expressed by the vector potential
of the laser field: $\mathbf{A}_{L}\mathbf{=-}c^{-1}\int_{0}^{t}\mathbf{E}%
\left( t^{\prime }\right) dt^{\prime }$ ($c$ is the light speed in vacuum).
Throughout this paper, for compactness of equations we will use the notation 
$\mathbf{A=}e\mathbf{A}_{L}/c\hbar $ for the vector potential. In the new
variables Eqs. (\ref{B1}) and (\ref{B3}) read: 
\begin{equation*}
\partial _{t}\mathcal{N}_{c}(\mathbf{k}_{0},t)=-2\mathrm{Im}\left\{ \left[
\hbar ^{-1}\mathbf{E}\left( t\right) \mathbf{D}_{\mathrm{tr}}\left( \mathbf{k%
}_{0}+\mathbf{A}\right) \right. \right.
\end{equation*}%
\begin{equation}
\left. \left. +\Omega _{c}\left( \mathbf{k}_{0}+\mathbf{A},t;\mathcal{P},%
\mathcal{N}_{c}\right) \right] \mathcal{P}^{\ast }(\mathbf{k}_{0},t)\right\}
,  \label{Bh1}
\end{equation}%
\begin{equation*}
\partial _{t}\mathcal{P}(\mathbf{k}_{0},t)=-i\hbar ^{-1}\left[ \mathcal{E}%
_{eh}\left( \mathbf{k}_{0}+\mathbf{A}\right) -i\hbar \Gamma \right] \mathcal{%
P}(\mathbf{k}_{0},t)
\end{equation*}%
\begin{equation*}
+i\left[ \hbar ^{-1}\mathbf{E}\left( t\right) \mathbf{D}_{\mathrm{tr}}\left( 
\mathbf{k}_{0}+\mathbf{A}\right) +\Omega _{c}\left( \mathbf{k}_{0}+\mathbf{A}%
,t;\mathcal{P},\mathcal{N}_{c}\right) \right]
\end{equation*}%
\begin{equation}
\times \left[ 1-2\mathcal{N}_{c}(\mathbf{k}_{0},t)\right] ,
\label{Bh2}
\end{equation}%
where $\mathcal{E}_{eh}\left( \mathbf{k}\right) =\mathcal{E}_{c}\left( 
\mathbf{k}\right) -\mathcal{E}_{v}\left( \mathbf{k}\right) -\Xi _{c}(\mathbf{%
k},t;\mathcal{P},\mathcal{N}_{c})$ is the electron-hole energy.  This set of
equations is equivalent to Eqs. (\ref{B1}) and (\ref{B3}), since the
gradient term $\mathbf{E}\left( t\right) \partial _{\mathbf{k}}$\ and the
time-dependent crystal momentum $\mathbf{k}_{0}+\mathbf{A}\left( t\right) $\
in Eqs. (\ref{Bh1}) and (\ref{Bh2}) describe the same effect. Without
Coulomb terms Eqs. (\ref{Bh1}) and (\ref{Bh2}) are semiconductor-Bloch
equation\ within the Houston basis \cite{Houston,SBE}. The surface current Eq. (
\ref{sfc}) can be split into the interband and intraband parts as follow: 
\begin{equation}
\mathbf{j}_{e}\left( t\right) =-\frac{2e}{(2\pi )^{2}}\int_{\widetilde{BZ}}d%
\mathbf{k}_{0}\left[ \mathbf{v}_{\mathrm{tr}}^{\ast }\left( \mathbf{k}_{0}+%
\mathbf{A}\right) \mathcal{P}(\mathbf{k}_{0},t)+\mathrm{c.c}\right] ,
\label{sfce}
\end{equation}%
\begin{equation}
\mathbf{j}_{a}\left( t\right) =-\frac{2e}{(2\pi )^{2}}\int_{\widetilde{BZ}}d%
\mathbf{k}_{0}\left[ \mathbf{v}_{c}\left( \mathbf{k}_{0}+\mathbf{A}\right) 
\mathcal{N}_{c}\left( \mathbf{k}_{0},t\right) +\mathrm{c.c}\right] ,
\label{sfca}
\end{equation}%
where $\mathbf{v}_{\mathrm{tr}}\left( \mathbf{k}\right) =i\left( \mathcal{E}%
_{c}\left( \mathbf{k}\right) -\mathcal{E}_{v}\left( \mathbf{k}\right)
\right) \mathbf{D}_{\mathrm{tr}}\left( \mathbf{k}\right) /e\hbar \ $is the
transition matrix element for velocity and $\mathbf{v}_{c}\left( \mathbf{k}%
\right) =\hbar ^{-1}\partial \mathcal{E}_{c}\left( \mathbf{k}\right)
/\partial \mathbf{k}\ $is the mean velocity of the conduction band. In Eqs. (%
\ref{sfce}) and (\ref{sfca}) we have taken into account electron-hole
symmetry and the BZ is also shifted to $\widetilde{BZ}=BZ-\mathbf{A}$.

The electron-electron interaction potential is modelled by screened Coulomb
potential \cite{Knorr2}: 
\begin{equation}
V_{2D}\left( \mathbf{q}\right) =\frac{2\pi e^{2}}{\epsilon \epsilon _{%
\mathbf{q}}\left\vert \mathbf{q}\right\vert },  \label{vc}
\end{equation}%
which accounts for the substrate-induced screening in the 2D nanostructure ($%
\epsilon $) and the screening stemming from other valence electrons ($%
\epsilon _{\mathbf{q}}$). Here, $\epsilon \equiv \left( \epsilon
_{1}+\epsilon _{2}\right) /2$, with the dielectric constants of the above $%
\epsilon _{1}$ and below $\epsilon _{2}$ surrounding media. Assuming a
graphene layer on a SiO$_{2}$ substrate ($\epsilon _{1}=1$, $\epsilon
_{2}=3.9$), we approximate $\epsilon =2.45$. The screening induced by
graphene valence electrons is calculated within the Lindhard approximation
of the dielectric function $\epsilon _{\mathbf{q}}$, which in a static limit
reads: $\triangleq $%
\begin{equation*}
\epsilon _{\mathbf{q}}=1-\frac{4\pi e^{2}}{\epsilon \left\vert \mathbf{q}%
\right\vert }
\end{equation*}%
\begin{equation}
\times \sum_{\lambda ,\lambda ^{\prime },\mathbf{k}}\digamma _{\lambda
,\lambda ^{\prime }}\left( \mathbf{k,q}\right) \frac{\mathcal{N}_{\lambda
^{\prime }}^{0}\left( \mathbf{k-q}\right) -\mathcal{N}_{\lambda }^{0}\left( 
\mathbf{k}\right) }{\mathcal{E}_{\lambda ^{\prime }}\left( \mathbf{k-q}%
\right) -\mathcal{E}_{\lambda }\left( \mathbf{k}\right) },  \label{DS}
\end{equation}%
where $\digamma _{\lambda \lambda ^{\prime }}\left( \mathbf{k,q}\right) =%
\frac{1}{2}\left( 1+\lambda ^{\prime }\lambda f_{c}\left( \mathbf{k,k}-%
\mathbf{q}\right) \right) $ is the band overlap function.

\begin{figure}[tbp]
\includegraphics[width=.5\textwidth]{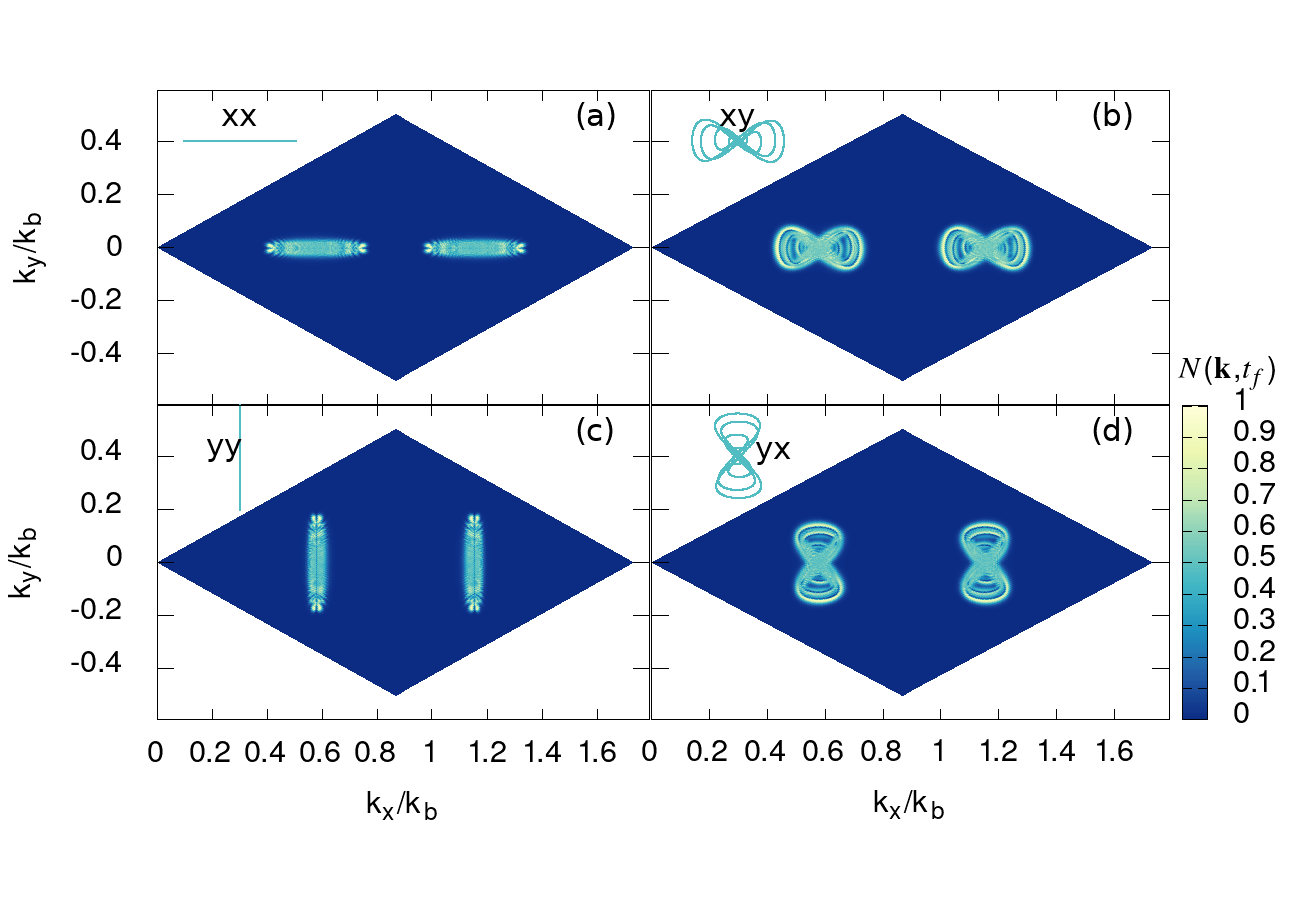}
\caption{ Particle distribution function $\mathcal{N}_{c}\left( \mathbf{k}%
,t_{f}\right) $ (in arbitrary units) after the interaction for graphene, as
a function of scaled dimensionless momentum components ($k_{x}/k_{b}$, $%
k_{y}/k_{b}$) for different relative polarizations of the fields. The
fundamental frequency is $\protect\omega =0.1\ \mathrm{eV}/\hbar $ and the
field strength is taken to be $E_{01}=E_{02}=4\ \mathrm{MV/cm}$ (intensity $%
2.\,13\times 10^{10}\mathrm{W\ cm}^{-2}$).}
\end{figure}

\begin{figure}[tbp]
\includegraphics[width=.45\textwidth]{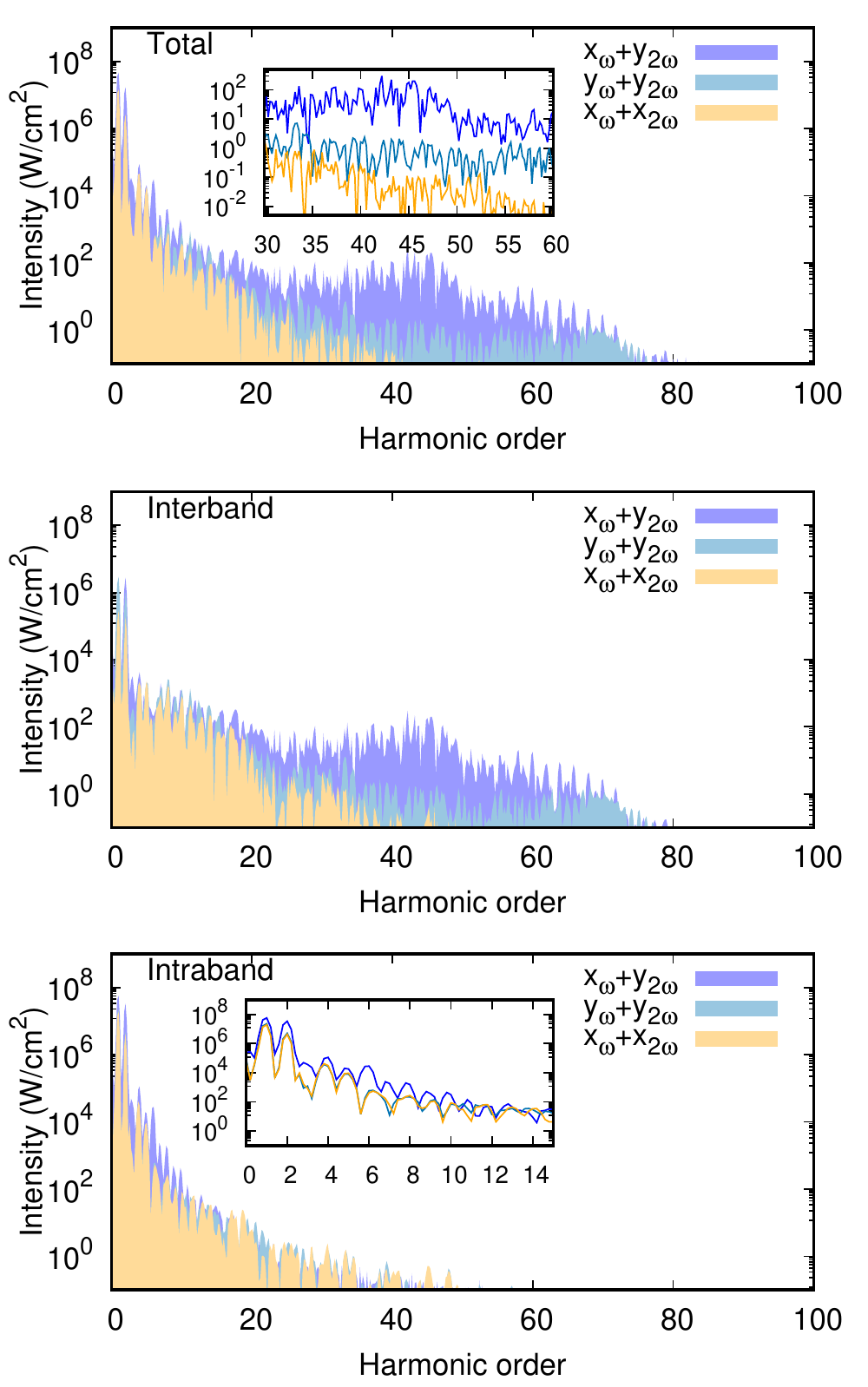}
\caption{The HHG spectra in logarithmic scale for graphene in the
strong-field regime for different relative polarizations of fundamental and
SH pump fields. The insets show the middle and the beginning of the spectra.
The fundamental frequency is $\protect\omega =0.1\ \mathrm{eV}/\hbar $ and
the field strength is taken to be $E_{01}=E_{02}=4\ \mathrm{MV/cm}$. }
\end{figure}

\section{Results}

In this section we discuss full numerical solution of the quantum HF equations 
(\ref{subsec:Quantum}) and to get more analytical insight, we study which effects 
can be already observed in a semiclassical approach (\ref{subsec:Semi}).

\subsection{\label{subsec:Quantum}Fully Quantum Calculations}

We explore the nonlinear response of graphene in a two-color laser field of
ultrashort duration. We take a 5-cycle fundamental laser field. The
envelopes and amplitudes of fundamental and SH fields are taken to be the
same: $E_{0}\equiv E_{01}=E_{02}$ and $\tau \equiv \tau _{1}=\tau _{2}$. The
amplitude of fundamental field was varied up to $8\ \mathrm{MV/cm}$
(intensity $8.\,\allowbreak 5\times 10^{10}\ \mathrm{W/cm}^{2}$). Hence, the
maximal intensity $1.\,\allowbreak 7\times 10^{11}\ \mathrm{W/cm}^{2}$
impending on graphene is below the damage threshold for monolayer graphene 
\cite{Yoshikawa,Roberts,Heide}. For all calculations, the relaxation time is
taken to be equal to the wave period $\Gamma ^{-1}=T=2\pi /\omega $. For the
considered frequencies we will have $\Gamma ^{-1}=20-40\ \mathrm{fs}$, which
is close to the experimental data \cite{Knorr1}. The combined laser field
possesses $C_{1}$ symmetry \cite{SR}, hence the allowed harmonic orders are $%
n\pm 1$, i.e. we have both even and odd harmonics. For sufficiently large 2D
sample the generated electric field far from the graphene layer is
proportional to the surface current: $\mathbf{E}^{(g)}(t)=-2\pi \lbrack 
\mathbf{j}_{e}(t)+\mathbf{j}_{a}(t)]/c$. The HHG spectral intensity is
calculated from the Fourier transform of the generated field.

\begin{figure}[tbp]
\includegraphics[width=.45\textwidth]{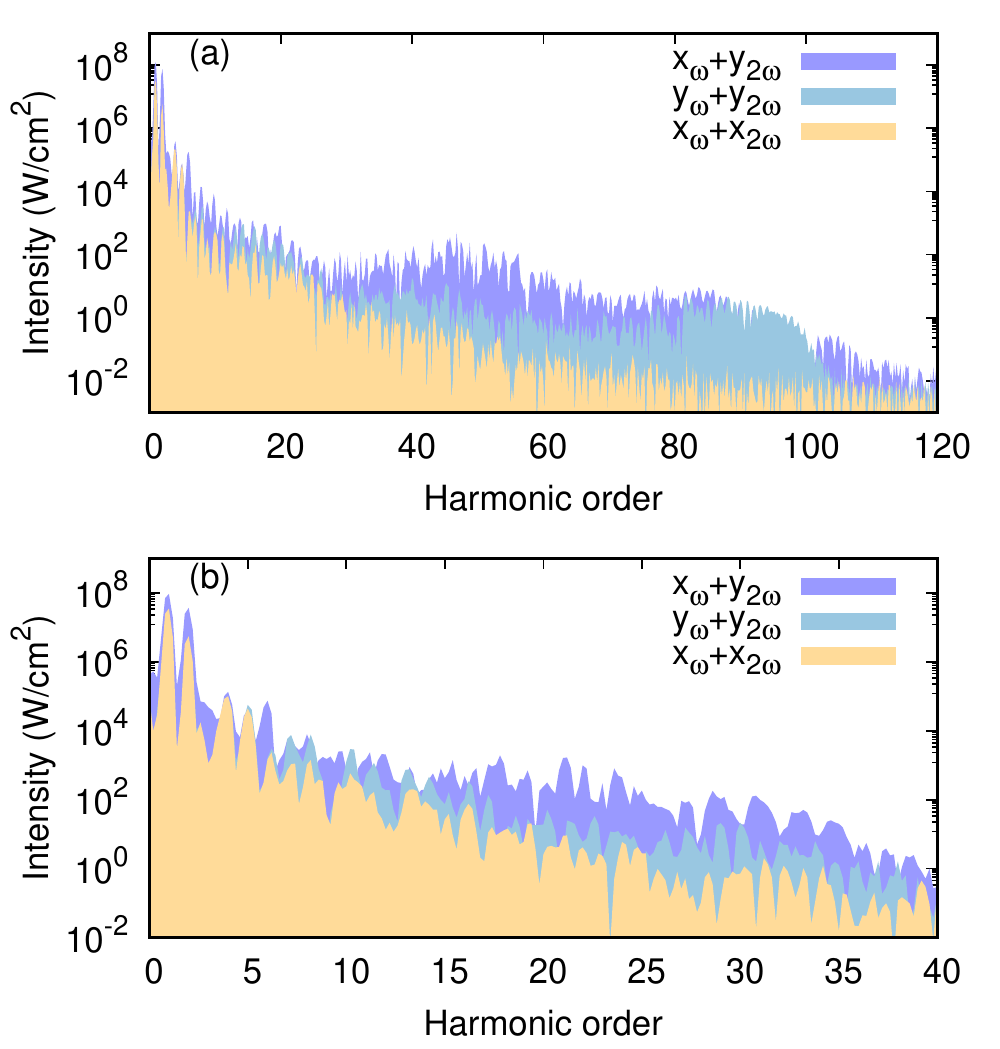}
\caption{The HHG spectra for graphene in the strong-field regime for
different relative polarizations of fundamental and SH pump fields. (a) The
fundamental frequency is $\protect\omega =0.1\ \mathrm{eV}/\hbar $ and the
field strength is taken to be $E_{01}=E_{02}=6\ \mathrm{MV/cm}$. (b) The
fundamental frequency is $\protect\omega =0.2\ \mathrm{eV}/\hbar $ and the
field strength is taken to be $E_{01}=E_{02}=8\ \mathrm{MV/cm}$. }
\end{figure}

\begin{figure*}[tbp]
\includegraphics[width=0.98\textwidth]{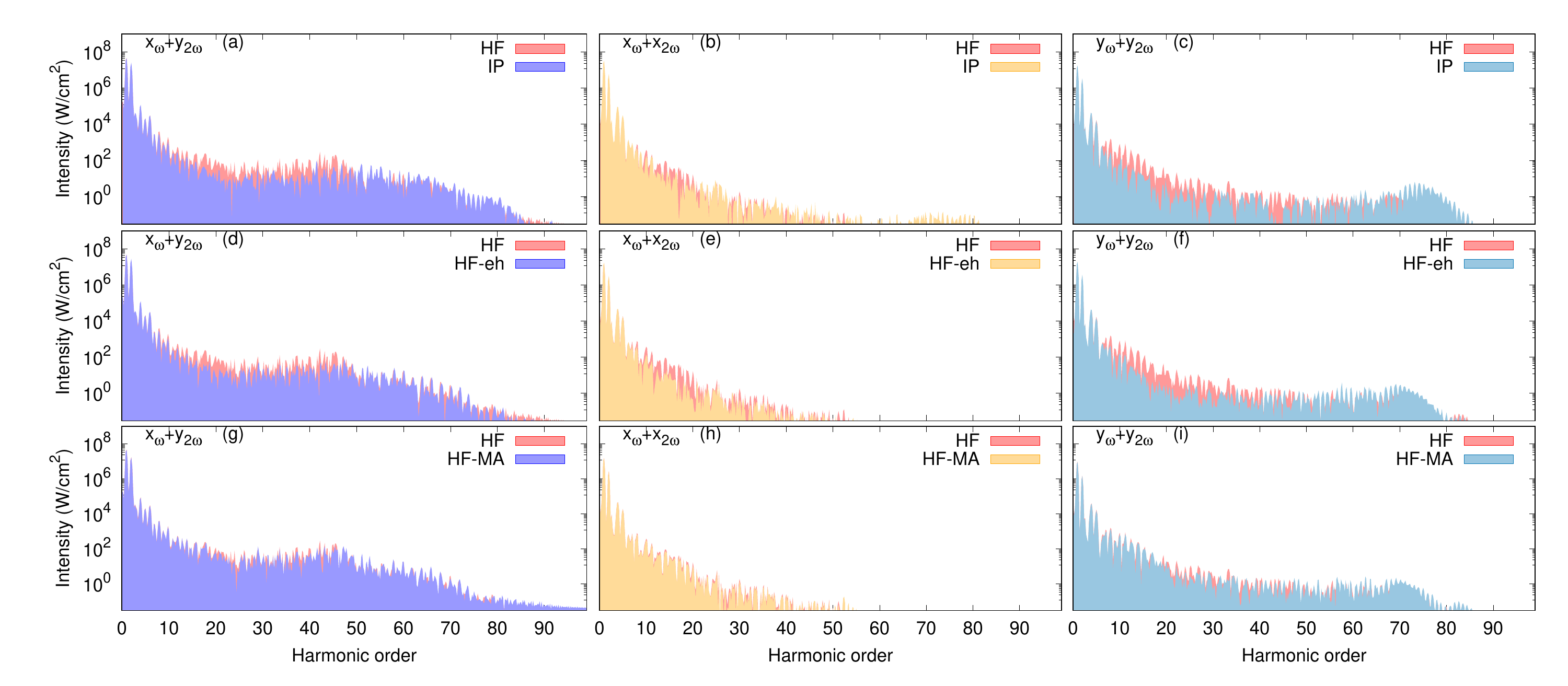}
\caption{The influence of the Coulomb interaction on the HHG process. The
fundamental frequency is $\protect\omega =0.1\ \mathrm{eV}/\hbar $ and the
field strength is taken to be $E_{01}=E_{02}=4\ \mathrm{MV/cm}$. (a-c)
comparison of $\protect\omega +2\protect\omega $ frequency mixing signals
for graphene with independent particles (IP) and with screened Coulomb
interaction in the HF level for different relative polarizations of
fundamental and SH pump fields. (d-f) comparison of full HF results with HF
approximation when only the repulsive part of the electron-electron
interaction is switched on. (g-i) comparison of full HF results with HF
approximation when MA processes are switched off.}
\end{figure*}
Thus, we have a set of nonlinear integro-differential equations (\ref{Bh1})
and (\ref{Bh2}), which have been solved numerically. The time propagation is
considered via fourth-order Runge-Kutta method. The numerically demanding
parts at each time step are the Coulomb contributions (\ref{Rc}) and (\ref%
{Wc}), which are integrated over the full BZ divided to $\sim $10$^{5}$
parts. These terms are calculated using the convolution theorem along with
the fast Fourier transform. This procedure decreases computational time by
two orders compared with the direct evolution. For integration, we take
rhombic BZ, where $\Gamma -K$ line is the x-axis. For the convergence of the
results we take $400\times 400$\ $\mathbf{k}$-points running parallel to the
reciprocal lattice vectors (\ref{basisb}).

In Fig. 2, we depict excitation of Fermi-Dirac sea, i.e. the electron distribution
function $\mathcal{N}_{c}\left( \mathbf{k},t_{f}=\tau \right) $ after the
interaction for graphene, as a function of scaled dimensionless momentum
components for different relative polarizations of the fields. In the left
corner of each density plot, we also show the Lissajous figures of
corresponding vector potentials $\mathbf{A}$. It is clearly seen that the
excitation patterns in the Fermi-Dirac sea follow the Lissajous diagrams.
The latter is the consequence of Eqs. (\ref{Bh1}) and (\ref{Bh2}). As is
seen for the orthogonal fields $x_{\omega }+y_{2\omega }$ and $y_{\omega }+x_{2\omega }$, we
have eight-like excitation shapes, while in the two parallel polarized fields 
$x_{\omega }+x_{2\omega }$ and $y_{\omega }+y_{2\omega }$ --
cigar-shape figures. As a consequence, in former cases the excitation areas
are considerably larger.

In Fig. 3, the HHG spectra in logarithmic scale with $\omega +2\omega $
frequency mixing for graphene in the strong-field regime for different
relative polarizations of fundamental and SH pump fields is presented. From
top to bottom we show the total, interband, and intraband parts of spectra.
The insets show the fine structure of HHG in the middle and in the beginning
of the spectra. As is seen from Fig. 3, in the case of the
orthogonally polarized two-color field the generated high-harmonics are
stronger than those obtained in the parallel polarization case by more than
two orders of magnitude. Such enhancement is colossal mainly for the
interband part of HHG which is predominant for the plateau part of the
spectrum. For the beginning of the spectrum where the intraband current is
dominant, we also have differences but not so noticeable. This tendency is
preserved also for the higher carrier frequency and intensity of laser pulses. This is
seen in Fig. 4, where we plotted the results of our calculations for larger
field strength -- Fig. 4(a) and for larger carrier frequency -- Fig. 4(b) compared
with Fig. 3. Note that our finding is in sharp contrast with the atomic \cite%
{Atom1} and semiconductor cases \cite{Navarrete} where the parallel
polarization case is more preferable. The reason of such discrepancy is the
vanishing gap for graphene, which ensures efficient creation of
electron-hole pairs with a large crystal momentum (see Fig. 2(b) and 2(d))
in case of orthogonally polarized two-color field. The non-zero crystal
momentum components eventually lead to re-encounter and annihilation of
these pairs after the acceleration in the laser fields with the subsequent
emission of high harmonics.

The polarization of harmonics depends on the symmetries of
the mean velocity $\mathbf{v}_{c}\left( \mathbf{k}\right) $ Eq. (\ref{Vk})
and the transition matrix element $\mathbf{v}_{\mathrm{tr}}\left( \mathbf{k}
\right) $ which is determined by the transition dipole moment $\mathbf{D}_{
\mathrm{tr}}\left( \mathbf{k}\right) $ (see Eq. (\ref{Dc})). In the parallel
polarization cases the polarizations of harmonics are evident. In Figs. (3) and (4) for the $
x_{\omega }+y_{2\omega }$ case the plateau harmonics are predominantly
polarized along the\textrm{\ }$y$\textrm{\ }axis, since $D_{x\mathrm{tr}
}\left( k_{x},-k_{y}\right) =-D_{x\mathrm{tr}}\left( k_{x},-k_{y}\right) $,
meanwhile excitation pattern is almost symmetrical with respect to the $
k_{y} $ axis, cf. Fig. 2(b).

Now, we will investigate the influence of the Coulomb interaction on the HHG
in graphene. Due to the vanishing bandgap, the screening is expected to be
large. On the other hand, the Coulomb interaction is known to be generally
stronger in low-dimensional structures. The comparison of $\omega +2\omega $
frequency mixing signals for graphene with independent charged carriers and
with Coulomb interaction in the HF level for different relative
polarizations of fundamental and SH pump fields is presented in the upper
three panels of Fig. 5. As is seen from these figures, in all cases we have
an overall enhancement of HHG signal due to Coulomb interaction. To obtain a
better insight, we also made calculations when electron-hole interaction is
switched off in Eqs. (\ref{Bh1}) and (\ref{Bh2}). The results in comparison
with HF approximation when only the repulsive part of the electron-electron
interaction is switched on are presented in Fig. 5(d-f). The light-matter
coupling (\ref{Rc}) and the transition energies (\ref{Wc}) contain
contributions from the population and the polarization via the MA processes.
For the systems with a vanishing gap, as in graphene, MA processes will be
essential especially for anisotropic excitation of the Fermi-Dirac sea. To
show the contribution of MA processes, in Figs. 5(g-i) we present the
results of calculations when the terms describing MA processes are switched
off in Eqs. (\ref{Bh1}) and (\ref{Bh2}). As is seen, in a linear scale, MA processes have
sizeable contributions in the enhancement of the HHG yield in the middle
part of the spectrum where the interband current is dominant. From Figs.
5(a-i) we conclude that the electron-hole interaction is responsible for the
enhancement of the interband HHG signal by almost one order of magnitude
compared with the independent charged carriers. Since this enhancement takes
place for all polarizations of driving waves, we can also conclude that
Coulomb interaction is not the reason for the enhancement of the HHG signal
stemming from the orthogonal polarization of two-color laser fields.

To clarify the observed enhancement further, in particular with respect to 
the vanishing band gap, we made a parallel consideration with gaseous and
semiconductor HHG and investigated $\omega +2\omega $ wave mixing for
graphene with artificially constructed energy gap (gapped graphene). For the
gap energy, we take $\Delta _{g}=2\ \mathrm{eV}$. The Coulomb interaction is
switched-off to avoid excitonic effects. In Fig. 6, we plot the HHG spectra
with $\omega +2\omega $ wave mixing for gapped graphene in the strong-field
regime at different relative polarizations. The laser parameters in Figs.
6(a) and 6(b) are the same as in Figs. 3 and 4(b), respectively. 
\begin{figure}[tbp]
\includegraphics[width=.45\textwidth]{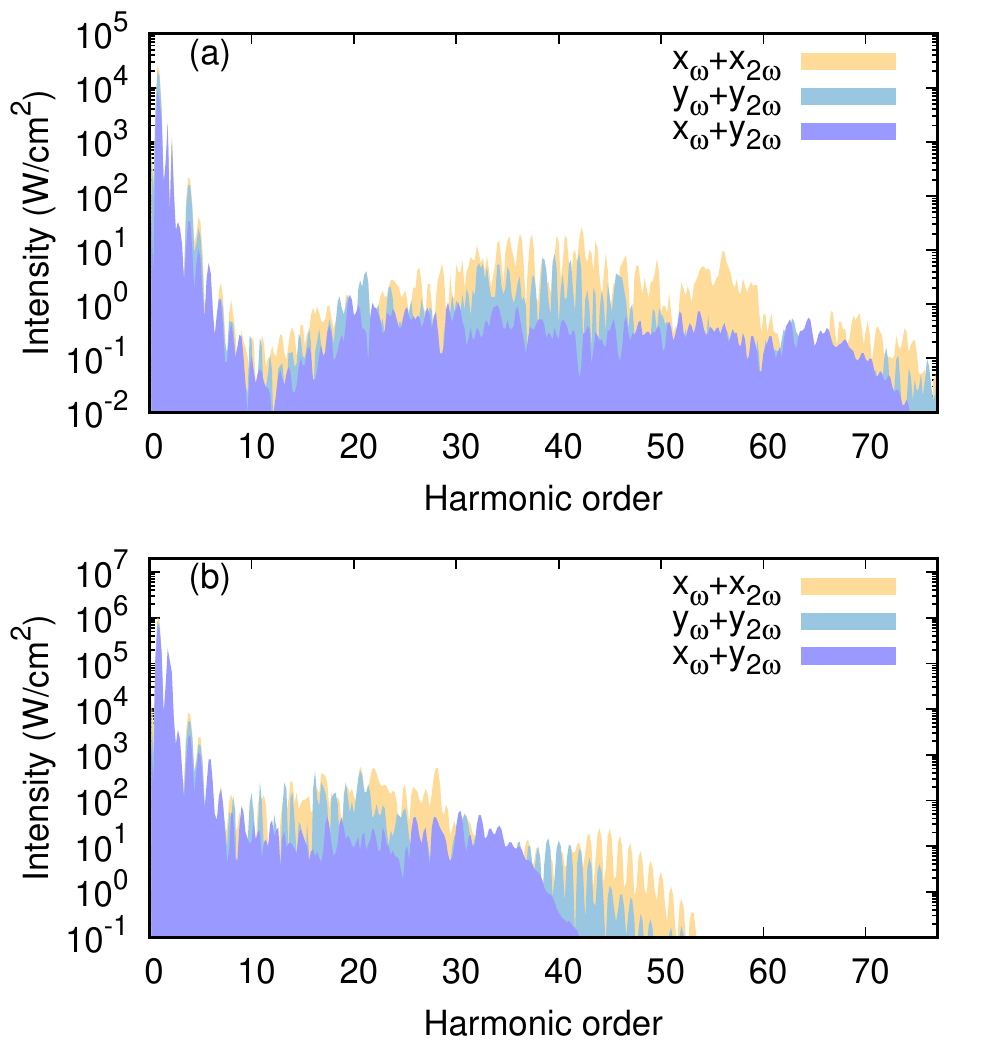}
\caption{The HHG spectra with $\protect\omega +2\protect\omega $ frequency
mixing for gapped graphene with $\Delta _{g}=2\ \mathrm{eV}$ in the
strong-field regime for different relative polarizations of fundamental and
SH pump fields. The electron-electron interaction is switched-off to avoid
excitonic effects. (a) The fundamental frequency is $\protect\omega =0.1\ 
\mathrm{eV}/\hbar $ and the field strength is taken to be $E_{01}=E_{02}=4\ 
\mathrm{MV/cm}$. (b) The fundamental frequency is $\protect\omega =0.2\ 
\mathrm{eV}/\hbar $ and the field strength is taken to be $E_{01}=E_{02}=8\ 
\mathrm{MV/cm}$.}
\end{figure}
Comparing Fig. 6 with Figs. 3 and 4 we see that the general picture in
considering regime for HHG is reverse, i.e. it is more preferable the
parallel polarization case. For the atomic HHG this is explained by the
three-step semiclassical recollision model. For gapped solid state system,
HHG is similar to atomic one \cite{Vampa2015} and can be modeled based on
the classical trajectory analysis of electron-hole pairs~\ref{subsec:Semi}. In this model
interband HHG occurs through the laser induced tunneling or multiphoton
creation of electron-hole pairs, which are accelerated in the laser field.
When pairs re-encounter, they recombine and a harmonic photon is created. If
the gap is considerably larger $\Delta _{g}>>\hbar \omega $ than the pump
wave photon energy the tunneling is the main mechanism for creation of
electron-hole pairs. Thus, electron-hole pairs are created near the minimum
of the electron-hole energy $\mathcal{E}_{eh}\left( \mathbf{k}\right) $,
i.e. near the Dirac points. For this case the particle distribution function 
$\mathcal{N}_{c}\left( \mathbf{k},t_{f}\right) $ after the interaction is
shown in Fig. 7. As is expected, the creation of electron-hole pairs out of
the Dirac points is exponentially suppressed, and the main contribution of
electron-hole pairs are Dirac points, which correspond to pairs with approximately
zero energy. It is also obvious that the tunneling probability will be
larger for parallel polarization case. Besides, it is also clear that for
orthogonally polarized two-color field to have re-encountering trajectories
one needs electron-hole pairs created with both non-zero crystal momentum
components, i.e. out of the Dirac points. The latter is suppressed for the
gapped system. For graphene, due to the vanishing bandgap, all these
conclusions do not hold, since the electron-hole pairs are created with
nonzero energy and the main question remains: can we extend the
semiclassical collision model for the considered case? The next subsection
will be devoted to investigation of this problem. 
\begin{figure}[tbp]
\includegraphics[width=.5\textwidth]{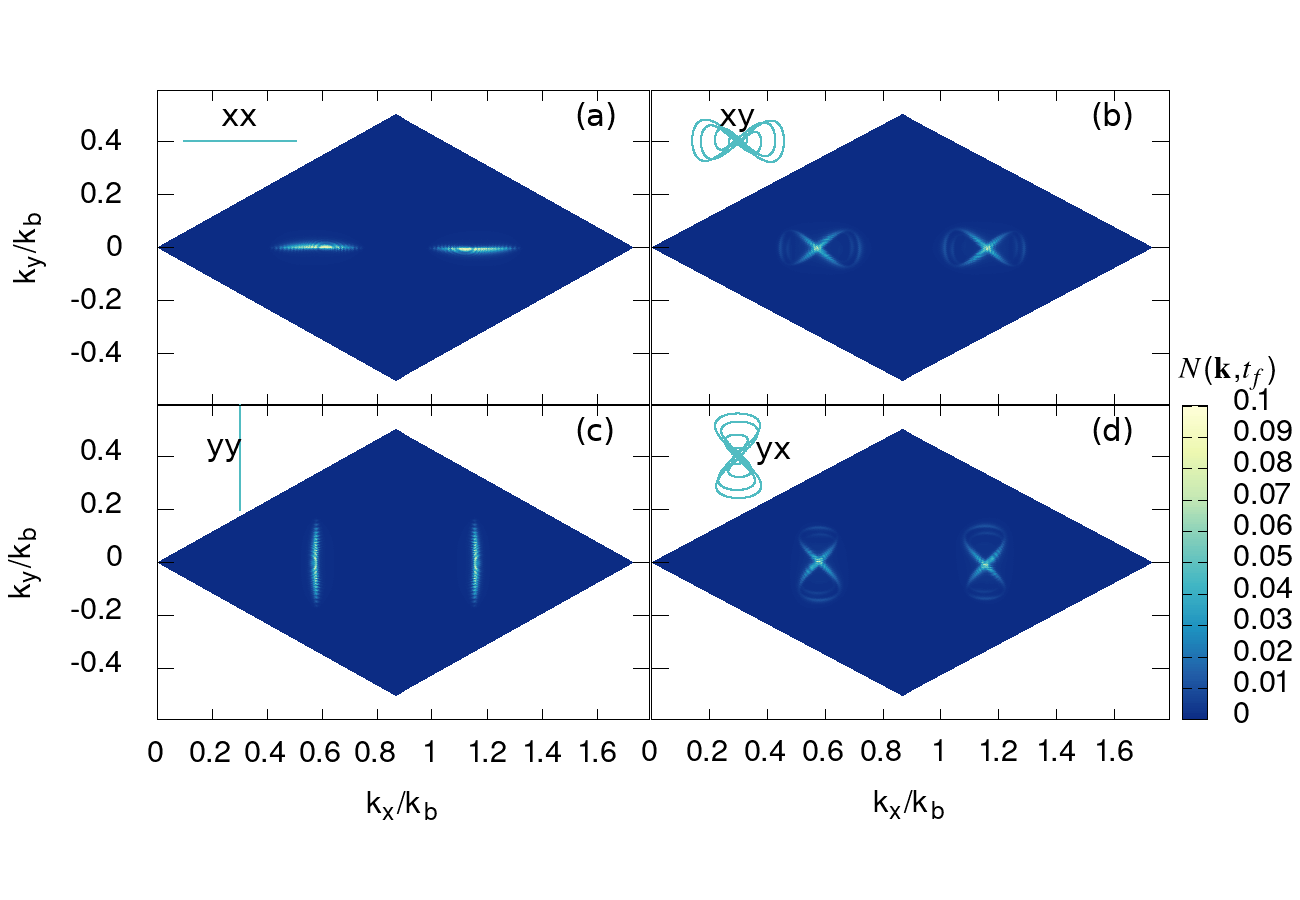}
\caption{ Particle distribution function $\mathcal{N}_{c}\left( \mathbf{k}%
,t_{f}\right) $ after the interaction for gapped graphene with $\Delta
_{g}=2\ \mathrm{eV}$, as a function of scaled dimensionless momentum
components ($k_{x}/k_{b}$, $k_{y}/k_{b}$) for different relative
polarizations of the fields. The electron-electron interaction is
switched-off to avoid excitonic effects. The remaining parameters coresponds
to the that of graphene.}
\end{figure}
\begin{figure*}[tbp]
\includegraphics[width=.99\textwidth]{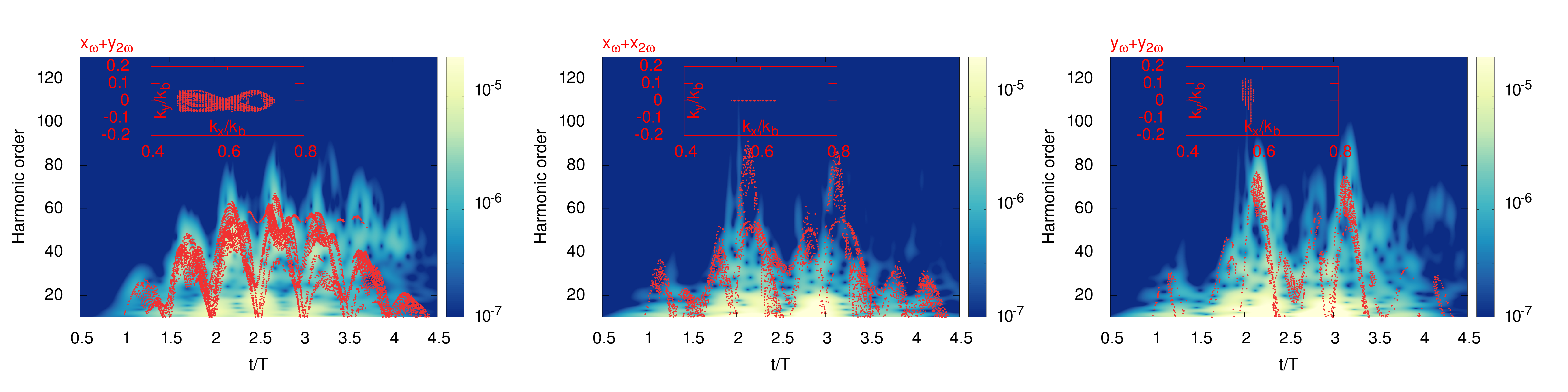}
\caption{The spectrogram of the HHG process via the wavelet transform of the
interband part of the surface current $\left\vert \mathbf{J}\left( t,\protect%
\omega \right) \right\vert $ (colorbox in arbitrary units). The laser
parameters are the same as in Fig. 3.  The red dots over the spectrograms are
the annihilation energies versus recollision times corresponding to the
solutions of saddle point equations (\protect\ref{sp1}), (\protect\ref{sp2}
), and (\protect\ref{sp3}). The insets for each case show the solutions of
the saddle-point equations for crystal momenta obtained with the same
resolution for half of the BZ.}
\end{figure*}

\subsection{\label{subsec:Semi}Semiclassical Collision Model}

To this end we need the evolution of high harmonic spectrum as a function of
time, we perform the time-frequency analysis by means of the wavelet
transform \cite{wavelet} of the interband part of the surface current:%
\begin{equation}
J_{\alpha }\left( t,\overline{\omega }\right) =\sqrt{\frac{\overline{\omega }%
}{\sigma }}\int_{0}^{\tau }dt^{\prime }j_{e,\alpha }\left( t^{\prime
}\right) e^{i\overline{\omega }\left( t^{\prime }-t\right) }e^{-\frac{%
\overline{\omega }^{2}}{2\sigma ^{2}}\left( t^{\prime }-t\right) ^{2}}.
\label{wavelet}
\end{equation}%
We have chosen Morlet wavelet with $\sigma =4\pi $. The spectrogram of the
HHG process via the wavelet transform of the interband part of the surface
current $\left\vert \mathbf{J}\left( t,\overline{\omega }\right) \right\vert $ is shown
in Fig. 8. The laser parameters are the same as in Fig. 3. In Fig. 8, for $%
x_{\omega }+y_{2\omega }$ case we have HHG emission peaks with period $0.5T$
starting at $t\approx 1.25T$ and ending at $4.25T$. These peaks coincide
with the peaks of the vector potential, i.e. when the absolute value of the
classical momentum given by the wave field is maximal. For the parallel
polarization cases, the peaks also are near the peaks of the vector
potential. This indicates that in some sense the above-obtained results can
also be understood in the scope of the semiclassical collision model \cite{Vampa2015}. For
this purpose, we now extend the semiclassical collision model to include
graphene without applying Keldysh approximation \cite{Keldysh64} at $%
\mathcal{N}_{c}\ll 1$. Thus, the formal non explicit solution of Eq. (\ref%
{Bh2}) for the interband polarization can be written as: 
\begin{equation}
\mathcal{P}(\mathbf{k}_{0},t)=\int_{0}^{t}dt^{\prime }e^{-\frac{i}{\hbar }%
S\left( \mathbf{k}_{0},t^{\prime },t\right) -\Gamma \left( t-t^{\prime
}\right) }\mathcal{K}^{eh}\left( \mathbf{k}_{0},t^{\prime }\right) ,
\label{pk}
\end{equation}%
where 
\begin{equation}
S\left( \mathbf{k}_{0},t^{\prime },t\right) =\int_{t^{\prime }}^{t}\left[ 
\mathcal{E}_{eh}\left( \mathbf{k}_{0}+\mathbf{A}\left( t^{\prime \prime
}\right) \right) \right] dt^{\prime \prime }  \label{clac}
\end{equation}%
is the classical action, and%
\begin{equation*}
\mathcal{K}^{eh}\left( \mathbf{k}_{0},t^{\prime }\right) =\frac{i}{\hbar }%
\left[ \mathbf{E}\left( t^{\prime }\right) \mathbf{D}_{\mathrm{tr}}\left( 
\mathbf{k}_{0}+\mathbf{A}\left( t^{\prime }\right) \right) +\hbar \Omega
_{c}\left( \mathbf{k}_{0}+\mathbf{A}\left( t^{\prime }\right) \right) \right]
\end{equation*}%
\begin{equation}
\times \left[ 1-2\mathcal{N}_{c}(\mathbf{k}_{0}+\mathbf{A}\left( t^{\prime
}\right) ,t^{\prime })\right]  \label{ehcr}
\end{equation}%
is the electron-hole creation amplitude. The latter is maximal near the
peaks of the total Rabi frequency and also includes the band filling factor $%
1-2N_{c}$, which reduces the electron-hole creation amplitude {due to Pauli
blocking. }Taking into account the definition (\ref{sfce}), the interband
part of the surface current can be represented in the following form:%
\begin{equation*}
j_{e,\alpha }\left( t\right) =-\frac{2}{(2\pi )^{2}}\int_{BZ}d\mathbf{k}%
_{0}\int_{0}^{t}dt^{\prime }\mathcal{K}^{eh}\left( \mathbf{k}_{0},t^{\prime
}\right)
\end{equation*}%
\begin{equation}
\times \exp \left[ -\frac{i}{\hbar }S\left( \mathbf{k}_{0},t^{\prime
},t\right) -\Gamma \left( t-t^{\prime }\right) \right] \mathcal{A}_{\alpha
}^{eh}\left( \mathbf{k}_{0},t\right) +\mathrm{c.c.},  \label{jle}
\end{equation}%
which have a transparent physical interpretation in analogy with the atomic
three-step model: electron-hole creation at $t^{\prime }$ with the amplitude 
$\mathcal{K}^{eh}\left( \mathbf{k}_{0},t^{\prime }\right) $, then
propagation in the $BZ$, which is defined by the classical action (\ref{clac}%
). At that, the propagation amplitude is diminished because of damping.
Finally, the electron-hole pair annihilates at $t$ with the amplitude 
\begin{equation}
\mathcal{A}_{\alpha }^{eh}\left( \mathbf{k}_{0},t\right) =\mathrm{v}_{%
\mathrm{tr},\alpha }\left( \mathbf{k}_{0}+\mathbf{A}\left( t\right) \right) .
\label{anha}
\end{equation}%
For the Fourier transform of the interband current, we will have%
\begin{equation*}
j_{e,\alpha }\left( \overline{\omega }\right) =-\frac{2}{(2\pi )^{2}}%
\int_{0}^{\tau }dt\int_{BZ}d\mathbf{k}_{0}\int_{0}^{t}dt^{\prime }e^{-\Gamma
\left( t-t^{\prime }\right) }\left[ \mathcal{K}^{eh}\left( \mathbf{k}%
_{0},t^{\prime }\right) \right.
\end{equation*}

\begin{equation}
\left. \times \exp \left[ -\frac{i}{\hbar }S\left( \mathbf{k}_{0},t^{\prime
},t\right) +i\overline{\omega }t\right] \mathcal{A}_{\alpha }^{eh}\left( 
\mathbf{k}_{0},t\right) +\mathrm{c.c.}\right] .  \label{jlew}
\end{equation}%
As is seen from this formula, the HHG can be driven by the nonlinearity of
the transition velocities by which the annihilation amplitude (\ref{anha})
is defined, or by the fast oscillatory part stemming from the classical
action. Assuming that the HHG is mainly driven by the last mechanism we can
further extend the collision model evaluating integrals in Eq. (\ref{jlew})
with the saddle point method. However, in contrast to the gapped system \cite%
{Keldysh64} one should relax the saddle point conditions for the following
reasons: since the gap is zero, the electron and hole can be created with
the initial energy ($\Delta _{\varepsilon }$) and due to the wave packet
spreading annihilation can take place at relative distance $\rho _{0}$.
Thus, we put the following conditions:%
\begin{equation*}
-\partial _{t^{\prime }}S<\Delta _{\varepsilon };\quad \left\vert \partial _{%
\mathbf{k}_{0}}S\right\vert <\rho _{0};\quad \partial _{t}S=\hbar \overline{%
\omega },
\end{equation*}%
which give 
\begin{equation}
\mathcal{E}_{eh}\left( \mathbf{k}_{0}+\mathbf{A}\left( t^{\prime }\right)
\right) <\Delta _{\varepsilon },  \label{sp1}
\end{equation}%
\begin{equation}
\left\vert \Delta \mathbf{r}\right\vert <\rho _{0,}  \label{sp2}
\end{equation}%
\begin{equation}
\mathcal{E}_{eh}\left( \mathbf{k}_{0}+\mathbf{A}\left( t\right) \right)
=\hbar \overline{\omega },  \label{sp3}
\end{equation}%
where 
\begin{equation}
\Delta \mathbf{r=}\int_{t^{\prime }}^{t}\left[ \mathbf{v}_{c}\left( \mathbf{k%
}_{0}+\mathbf{A}\left( t^{\prime \prime }\right) \right) -\mathbf{v}%
_{v}\left( \mathbf{k}_{0}+\mathbf{A}\left( t^{\prime \prime }\right) \right) %
\right] dt^{\prime \prime }  \label{dr}
\end{equation}%
is the electron-hole separation vector and $\mathbf{v}_{c,v}=\hbar
^{-1}\partial _{\mathbf{k}_{0}}\mathcal{E}_{c,v}\left( \mathbf{k}_{0}\right) 
$ are the group velocities. Saddle point equations (\ref{sp1}), (\ref{sp2}),
and (\ref{sp3}) have the following interpretation. The first one defines the
birth time ($t^{\prime }$) at which the electron-hole pair is formed. It
also states that the electron-hole pair is created with initial momentum
defined by the area of the excited Fermi-Dirac sea. According to the second
condition the laser accelerates the electron and hole with the instantaneous
group velocities $\mathbf{v}_{c,v}$ and depending on the creation time the
electron-hole may recollide at the time $t$ with final momentum $\mathbf{k}%
_{0}+\mathbf{A}\left( t\right) $ and relative distance $\rho _{0}$. The
third condition is the conservation of energy: the electron-hole annihilate,
emitting the energy in the form of a single photon. Taking into account Fig.
2, for $\Delta _{\varepsilon }$ we take $2\ $eV and $\rho _{0}=2a$.

\begin{figure}[tbp]
\includegraphics[width=.39\textwidth]{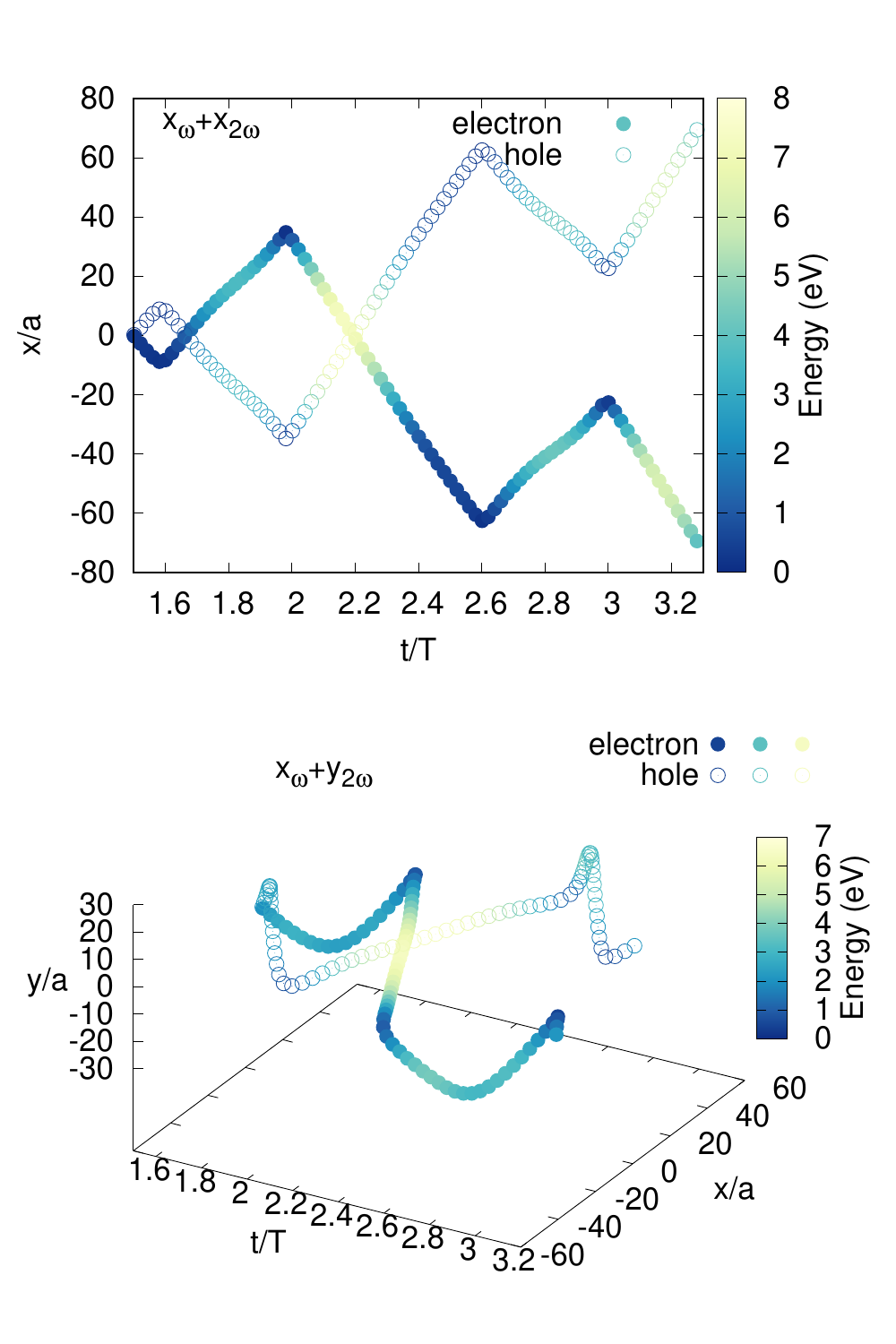}
\caption{The colliding trajectories of an electron and a hole. (a) The
electron (hole) is created at the $t_{c}=1.5T$ with the initial crystal
momentum $k_{x}/k_{b}=0.57$ (-0.57), $k_{y}/k_{b}=0$(0) which corresponds to
initial energy 0.3 eV. The colored trajectory and colored box show the
energies acquired by carriers along the trajectory. (b) The electron (hole)
is created at the $t_{c}=1.575T$ , with the initial crystal momentum $
k_{x}/k_{b}=0.6$ ($-0.6$), $k_{y}/k_{b}=-0.05$ (0.05) which corresponds to
initial energy 2 eV.}
\end{figure}
We then integrate the equation $\mathbf{r}_{e}\left( t^{\prime },t\right) 
\mathbf{=}\int_{t^{\prime }}^{t}\left[ \mathbf{v}_{c}\left( \mathbf{k}_{0}+%
\mathbf{A}\left( t^{\prime \prime }\right) \right) \right] dt^{\prime \prime
}$ to obtain the classical motion of the electron. Since we have
electron-hole symmetry, for the hole we have $\mathbf{r}_{h}\left( t^{\prime
},t\right) =-\mathbf{r}_{e}\left( t^{\prime },t\right) $. We also calculate
the electron-hole distance $\rho \left( t^{\prime },t\right) =\left\vert 
\mathbf{r}_{e}-\mathbf{r}_{h}\right\vert $. It is assumed the colliding
trajectories if at $t>t^{\prime }$ we have a local minimum of the
electron-hole distance $\rho _{m}\left( t^{\prime },t\right) <\rho _{0}$.
Then we fix the time and the corresponding energies $\mathcal{E}_{eh}\left( 
\mathbf{k}_{0}+\mathbf{A}\left( t\right) \right) $.

In Fig. 9 the typical colliding trajectories of an electron and a hole are
shown. In Fig. 9(a) both fields are in the x-direction and we have
one-dimensional motion. The electron-hole pair is created near the Dirac
point. The laser field accelerates the electron and hole with the
instantaneous group velocities $\mathbf{v}_{e}=-\mathbf{v}_{h}$. The colored
trajectory and colored box show the energies acquired by carriers along the
trajectory. We see two collisions: with the low energy $1.2\ \mathrm{eV}$
and the second one at $t=2.2T$ with the high energy $\mathcal{E}_{eh}\left( 
\mathbf{k}_{0}+\mathbf{A}\left( t\right) \right) /\hbar \omega \simeq 71$.
In Fig. 9(b) wave fields are perpendicular to each other. In this case for
the colliding trajectories, one needs electrons and holes created with both
non-zero crystal momentum components. In this case we have two-dimensional
motion. The electron-hole pair is created at the $t_{c}=1.575T$ with the
initial energy $2.0\ \mathrm{eV}$ . Then, after the laser acceleration, the
electron and hole collision/annihilation take place at $t_{a}=2.2T$. Here
the collision is not perfect because of two- dimensional motion. The minimal
distance is $0.6a$.

Varying the creation time $t^{\prime }$ from $0.5T$ to $4.5T$ we have solved
Eqs. (\ref{sp1}), (\ref{sp2}), and (\ref{sp3}). The corresponding
annihilation/recombination energies versus recollision times are plotted in
Fig. 8 over the spectrograms of the HHG process. As is seen from these
plots, we have a fairly good agreement with fully quantum calculations. 

The semiclassical approach has transparent physical interpretation in terms
of Feynman's path integral \cite{Sal}. An infinite-dimensional functional
integral is reduced to a coherent sum over the finite quantum trajectories
defined by the saddle point equations \cite{Vampa2014,Li-Lan}. This way, Eq.
(\ref{jlew}) can be approximated as 

\begin{equation*}
j_{e,\alpha }\left( \overline{\omega }\right) \propto \sum_{\mathbf{k}
_{0n},t_{n},t_{n}^{\prime }}A_{n}e^{-\Gamma \left( t_{n}-t_{n}^{\prime
}\right) }\left[ \mathcal{K}^{eh}\left( \mathbf{k}_{0n},t_{n}^{\prime
}\right) \right.
\end{equation*}

\begin{equation}
\left. \times \exp \left[ -\frac{i}{\hbar }S\left( \mathbf{k}
_{0n},t_{n}^{\prime },t_{n}\right) +i\overline{\omega }t_{n}\right] \mathcal{
A}_{\alpha }^{eh}\left( \mathbf{k}_{0n},t_{n}\right) +\mathrm{c.c.}\right] .
\label{FP}
\end{equation}
where $\mathbf{k}_{0n},t_{n},t_{n}^{\prime }$\ are the solutions of the
saddle point equations (\ref{sp1})-(\ref{sp3}), $A_{n}$\ is an amplitude and 
$S$ is defined in (\ref{clac}). One obtains qualitatively and quantitatively
different results depending on how many and which trajectories are involved
in Eq. (\ref{FP}). The insets in Fig. 8 for each case show the solutions of
the saddle-point equations for crystal momenta obtained with the same
resolution for half of BZ. Note that for each solution $\mathbf{k}_{0n}$
there may be several trajectories that can interfere constructively or
destructively in Eq. (\ref{FP}) depending on the phase factor. In the case
of an orthogonally polarized two-color field, the crystal momenta obeying
saddle point equations (\ref{sp1})-(\ref{sp3}) and, consequently, the
trajectories are an order of magnitude larger than in the case of parallel
polarization, cf. Fig. 8. At that, a large number of trajectories with the
different $\mathbf{k}_{0n}$ but the close $t_{n}^{\prime },t_{n}$
considerably enlarges the number of constructively interfering terms in Eq. (
\ref{FP}). For this reason, $j_{e,\alpha }\left( \overline{\omega }\right) $
enhances in an orthogonally polarized two-color field. For the gapped system
this favorable condition disappears. Thus, the saddle point equation (\ref
{sp1}) that defines the birth time ($t^{\prime }$) becomes $E_{eh}\left( 
\mathbf{k}_{0}+\mathbf{A}\left( t^{\prime }\right) \right) \approx -\Delta
_{g}$, which cannot be satisfied for any real time. Because of the imaginary
solution $t^{\prime }$, we have an exponential suppression of the
probability of electron and hole production, especially with momenta outside
the Dirac points. In addition, tunneling occurs near the maxima of the
electric field strength \cite{Keldysh64} and is more probable for parallel
polarization of a two-color field (\ref{Et}). These are the main reasons why
in the gapped nanostructure HHG in an orthogonally polarized two-color field
is suppressed compared with the case of parallel polarization. Also, note
that for the observation of the HHG yield in the orthogonally polarized
two-color field the incident intensity of SH wave should be comparable to or
larger than the incident fundamental wave intensity to ensure the existence
of re-encountering trajectories Eq. (\ref{sp2}), which maximize the
recombination probability.

\section{Conclusion}

We have presented the microscopic theory of nonlinear interaction of a
monolayer graphene with a strong bichromatic few-cycle driving pulse that is
composed of the superposition of an infrared fundamental pulse of linear
polarization and its second harmonic at the parallel and orthogonal
polarizations. The electron-electron Coulomb interaction has been taken into
account in the scope of HF approximation beyond the Dirac cone
approximation, which is applicable to the full Brillouin zone of the
hexagonal tight-binding nanostructure. The obtained results show that in all
cases we have an overall enhancement of HHG yield compared with the
independent charged carrier model due to the electron-hole attractive
interaction. We have shown that in the case of the orthogonally polarized
two-color field the generated high-harmonics are stronger than those
obtained in the parallel polarization case by more than two orders of
magnitude. This enhancement is colossal mainly for the interband part of HHG
that is predominant for the plateau part of the spectrum. This tendency
persists for a wide range of intensities and frequencies of driving waves.
The physical origin of polarization-dependent strong enhancement in graphene
is also deduced from the three-step semiclassical electron-hole collision
model, extended to graphene with pseudo-relativistic energy dispersion and
without Keldysh approximation.

\begin{acknowledgments}
The work was supported by the Science Committee of Republic of
Armenia (SCRA), project No. 21AG-1C014. AK (TU Berlin) thanks SCRA for supporting the exchange between
Yerevan State University and Technische Universit\"{a}t Berlin.
\end{acknowledgments}

\appendix

\section{The single-particle Hamiltonian and other derived quantities}

In this section we consider the details of the single particle Hamiltonian
and give concrete expressions for the used physical quantities. The single
particle Hamiltonian is taken to be:%
\begin{equation}
\widehat{H}=\left[ 
\begin{array}{cc}
0 & -\gamma _{0}f\left( \mathbf{k}\right) \\ 
-\gamma _{0}f^{\ast }\left( \mathbf{k}\right) & 0%
\end{array}%
\right] ,  \label{H(k)}
\end{equation}%
where $\gamma _{0}=2.8\ $eV. The structure function is 
\begin{equation}
f\left( \mathbf{k}\right) =e^{i\frac{ak_{y}}{\sqrt{3}}}+2e^{-i\frac{ak_{y}}{2%
\sqrt{3}}}\cos \left( \frac{ak_{x}}{2}\right) ,  \label{f(k)}
\end{equation}%
where $a$ is the lattice spacing. The reciprocal lattice unit cell is a
rhombus formed by \ \ two vectors:%
\begin{equation}
\mathbf{b}_{1}=\left( \frac{2\pi }{a},-\frac{2\pi }{a\sqrt{3}}\right) ,\quad 
\mathbf{b}_{2}=\left( \frac{2\pi }{a},\frac{2\pi }{a\sqrt{3}}\right) .
\label{basisb}
\end{equation}%
The low-energy excitations are centered around the two points $\mathrm{K}%
_{+} $ and $\mathrm{K}_{-}$represented by the vectors%
\begin{equation}
\mathbf{K}_{+}=\frac{k_{b}}{\sqrt{3}}\widehat{\mathbf{x}},\quad \mathbf{K}%
_{-}=\frac{2k_{b}}{\sqrt{3}}\widehat{\mathbf{x}},  \label{K+-}
\end{equation}%
where $k_{b}=4\pi /\sqrt{3}a$ . Note that near the two Dirac points $\gamma
_{0}f\left( \mathbf{k}\right) =i\mathrm{v}_{F}\hbar k_{y}\mp \mathrm{v}%
_{F}\hbar k_{x}$, where $\mathrm{v}_{F}=\sqrt{3}a\gamma _{0}/2\hbar $ is the
Fermi velocity. The eigenstates of the Hamiltonian (\ref{H(k)}) are%
\begin{equation}
|\lambda ,\mathbf{k}\rangle =\frac{1}{\sqrt{2}}\left[ 
\begin{array}{c}
-\frac{f\left( \mathbf{k}\right) }{\left\vert f\left( \mathbf{k}\right)
\right\vert } \\ 
\lambda%
\end{array}%
\right] ,  \label{spinor}
\end{equation}%
corresponding to energies%
\begin{equation}
\mathcal{E}_{\lambda }\left( \mathbf{k}\right) =\lambda \gamma
_{0}\left\vert f\left( \mathbf{k}\right) \right\vert .  \label{energy}
\end{equation}%
Here the band index $\lambda =\pm 1$ (for conduction ($\lambda =1$) and
valence ($\lambda =-1$) bands). The transition dipole moment is explicitly
given by%
\begin{equation*}
\mathbf{D}_{\mathrm{tr}}\left( \mathbf{k}\right) =-\frac{ea}{2\left\vert
f\left( \mathbf{k}\right) \right\vert ^{2}}\sin \left( \frac{\sqrt{3}}{2}%
ak_{y}\right) \sin \left( \frac{ak_{x}}{2}\right) \widehat{\mathbf{x}}
\end{equation*}%
\begin{equation}
+\frac{ea}{2\sqrt{3}\left\vert f\left( \mathbf{k}\right) \right\vert ^{2}}%
\left( \cos \left( ak_{x}\right) -\cos \left( \frac{\sqrt{3}}{2}%
ak_{y}\right) \cos \left( \frac{ak_{x}}{2}\right) \right) \widehat{\mathbf{y}%
}.  \label{Dc}
\end{equation}%
The band velocity is given by the formula%
\begin{equation*}
\mathbf{v}_{c}\left( \mathbf{k}\right) =-\mathrm{v}_{F}\frac{2}{\sqrt{3}%
\left\vert f\left( \mathbf{k}\right) \right\vert }
\end{equation*}%
\begin{equation*}
\times \left[ \cos \left( \frac{\sqrt{3}}{2}ak_{y}\right) \sin \left( \frac{%
ak_{x}}{2}\right) +\sin \left( ak_{x}\right) \right] \widehat{\mathbf{x}}
\end{equation*}%
\begin{equation}
-\mathrm{v}_{F}\frac{2}{\left\vert f\left( \mathbf{k}\right) \right\vert }%
\sin \left( \frac{\sqrt{3}}{2}ak_{y}\right) \cos \left( \frac{ak_{x}}{2}%
\right) \widehat{\mathbf{y}}.  \label{Vk}
\end{equation}%
For the gapped graphene we used general formulas \cite{Mer19} taking also
into account Berry connections.

\end{document}